\documentstyle[preprint,aps]{revtex}
\begin{document}
\draft
\title{Analytical calculation of matter effects in two mass-scale 
neutrino oscillations} 
\author
{ Mohan Narayan and S. Uma Sankar}
\address
{Department of Physics, Indian Institute of Technology, Bombay}
\date{\today}
\maketitle

\begin{abstract}
We consider three active flavor neutrino oscillations where both the
mass-square differences play a role in atmospheric neutrino problem.
We calculate the matter effects arising due to propagation through 
earth. We demonstrate that these effects improve the fit to the 
electron data {\it vis-a-vis} vacuum oscillations but make the fit to 
the muon data far worse, thus
worsening the overall fit. The results of our analytical calculation
verify the numerical investigations of this scheme presented earlier 
by Fogli {\it et al}.
\end{abstract}
%\vspace{0.5cm}
%\pacs{PACS numbers:  14.60.Gh, 96.60.Kx, 95.30.Cq, 96.40.Tv}
\narrowtext

\section{Introduction}
There are three evidences for neutrino oscillations: 1) solar neutrino
problem, 2) atmospheric neutrino problem and 3) the results of LSND 
experiment. Two flavor oscillation analysis of each individual problem 
gives three widely 
different values for neutrino mass-squared differences ($\Delta$'s).
Oscillations between the three active flavor neutrinos have only two
independent $\Delta$'s.
Hence it was proposed that there should be
at least one sterile neutrino if one wants to explain all three evidences
in terms of neutrino oscillations \cite{srubabati}. However, various 
attempts were made to account for all three evidences in terms of three
active flavor oscillations \cite{nru}. CHOOZ constraint on $13$ mixing
angle rules out most of these solutions \cite{chooz}. 

Recently three active flavor oscillations as explanation for all three  
evidences was revived with the following scenario \cite{sb,tm}. The larger 
$\Delta$ is
about $0.3$ eV$^2$ which can explain LSND result. The smaller $\Delta$
is in the range $10^{-4}-10^{-3}$ eV$^2$. The upper limit
is set so that CHOOZ constraint on the relevant mixing angles will not
apply to oscillations driven by this $\Delta$. The lower
limit is set by demanding that the solar neutrino survival probability  
should be independent of energy. 
In these three flavor oscillation schemes,
the atmospheric neutrino oscillations are driven by both the $\Delta$'s. 
In particular, for the oscillations driven by the smaller $\Delta$,
the oscillation probability is almost zero 
for downward going neutrinos and is significant for upward going neutrinos.
In particular these oscillations create the zenith angle dependence 
which is needed to explain the data. 

This scheme reproduces the overall suppression seen in the
atmospheric neutrino data and also the zenith dependence of 
the ratio of muon to electron events. But it does not give 
a flat dependence for the zenith distribution for electron 
events which is observed in the data. It also gives an incorrect
zenith distribution for muon events. The large $\nu_e 
\leftrightarrow \nu_\mu$ transitions driven by the smaller
$\Delta$, which are needed to explain solar
neutrino problem, are responsible for this distortion of 
zenith angle distribution in atmospheric neutrino problem.
In ref. \cite{fogli} this scheme was investigated numerically. 
Their results are summarized in figs (4) and (5) of \cite{fogli}. 
There it was also pointed out that matter effects worsen the already
bad fit to data. Three salient features emerge from these graphs.
\begin{itemize}
\item 
For both sub-GeV electron as well as muon events there is no appreciable
difference in the event distributions with and without matter. This implies  
that matter and vacuum oscillation probabilities are the same.
\item 
For multi-GeV electron events matter greatly suppresses the large $\nu_{\mu}
\leftrightarrow \nu_e$ oscillations and thus drastically reduces the large 
excesses of electron events which vacuum oscillations generate especially
at large distances of traversal, i.e for the upward going events.
\item 
Both for multi-GeV muon events and upgoing muon events inclusion of matter
effects results in essentially a 
flat profile as a function of the zenith angle.
\end{itemize}

In this paper we show that these features can be explained by a  
simple analytic calculation of matter effects in this scheme by using
perturbation theory. Very recently two scale oscillations of 
of atmospheric neutrinos were treated analytically \cite{cgk,bds}
under a different set of neutrino masses and mixings. But these 
analyses can explain only LSND results and atmospheric neutrino
data leaving solar neutrino problem untouched.

\section{Two scale oscillations in matter- a perturbative
analysis}

We take Scheck-Barenboim (SB) scheme given in \cite{sb} as 
a representative of models in which all evidences for neutrino
oscillations are explained in terms of three active flavors.
A recent update using latest reactor and 
accelerator data is given in \cite{bnu}.
The masses of the vacuum mass eigenstates are taken to be $\mu_1$,
$\mu_2$ and $\mu_3$.
The vacuum oscillation here is controlled by both the mass-squared
differences 
$\Delta_{21} = \mu_2^2 - \mu_1^2$ and $\Delta_{31} = \mu_3^2 -
\mu_2^1$. 
$\Delta_{31}$ is chosen to be about $0.3$ eV$^2$ to
drive the oscillations seen at LSND and $\Delta_{21}$ is constrained to be
a few times $10^{-4}$ eV$^2$. 
Both scales average out to give an energy independent suppression in the
solar neutrino case. The (12) mixing angle $\omega$ should be close to
$\pi/4$, so that the solar neutrino survival probability is $0.5$.
A recent analysis found that this energy independent solution of solar
neutrino problem is  allowed with good confidence level \cite{cggr,bgp}.
The (13) mixing angle $\phi$ 
is constrained to be quite small $\leq 10^\circ$
by reactor and LSND data. The (23) mixing angle $\psi$ is
centered around 27 degrees to give a good fit to the ratio of 
muon to electron events in atmospheric neutrino data.

The three flavor eigenstates
are related to the three mass eigenstates in vacuum through a unitary
transformation,
\begin{equation}
\left[ \begin{array}{c} \nu_e \\ \nu_{\mu} \\ \nu_{\tau}
\end{array} \right] = U^v
\left[ \begin{array}{c} \nu_1^v \\ \nu_2^v \\ \nu_3^v
\end{array} \right],
\end{equation}
where the superscript $v$ on r.h.s. stands for vacuum.
The $3 \times 3$ unitary matrix $U^v$ can be parametrized by three Euler
angles $(\omega, \phi, \psi)$ and a phase. The form of the
unitary matrix can therefore be written in general as,
$ U^{v} = U_{23}(\psi) \times U_{phase} \times U_{13}(\phi)\times
U_{12}(\omega),$
where $U_{ij}(\theta_{ij})$ is the mixing matrix between $i$th and $j$th
mass eigenstates with the mixing angle $\theta_{ij}$.
The explicit form of $U$ is
\begin{equation}
U^v = \left( \begin{array}{ccc}
 c_{\phi} c_{\omega} & c_{\phi} s_{\omega} & s_{\phi} \\
-c_{\psi} s_{\omega} e^{i \delta} - s_{\psi} s_{\phi} c_{\omega} e^{-i \delta} & 
c_{\psi} c_{\omega} e^{i \delta} - s_{\psi} s_{\phi} s_{\omega} e^{-i \delta} & 
  s_{\psi} c_{\phi} e^{-i \delta}   \\
s_{\psi} s_{\omega} e^{i \delta} - c_{\psi} s_{\phi} c_{\omega} e^{-i \delta} & 
-s_{\psi} c_{\omega} e^{i \delta} - c_{\psi} s_{\phi} s_{\omega} e^{-i \delta} & 
 c_{\psi} c_{\phi} e^{-i \delta}   
\end{array} \right), \label{eq:defUv}
\end{equation}
where $s_{\phi} = \sin \phi$ and $c_{\phi} = \cos \phi$ etc. All the angles 
can take values between $0$ and $\pi/2$.  
$U^{v}$  can also be written  as
\begin{equation}
U^v = \left( \begin{array}{ccc}
 U_{e 1} & U_{e 2} & U_{e 3}  \\
 U_{\mu 1} & U_{\mu 2} & U_{\mu 3}  \\
 U_{\tau 1} & U_{\tau 2} & U_{\tau 3} 
\end{array} \right) .
\end{equation}

In the mass eigenbasis, the $({\rm mass})^2$ matrix is diagonal,
\begin{eqnarray}
M_0^2  & = &  \left( \begin{array}{ccc}
			 0 & 0 & 0 \\
			 0 & \Delta_{21} & 0 \\
			 0 & 0 & \Delta_{31} \\
			 \end{array} \right),
\end{eqnarray}

In the flavour basis
the $({\rm mass})^2$ matrix has the form
\begin{eqnarray}
M_v^2  & = & U^v M_0^2 {U^v}^{\dagger} \nonumber \\
       & = & \Delta_{31} M_{31} + \Delta_{21} M_{21},
\end{eqnarray}
where
\begin{eqnarray}
M_{31} & = & U^v  \left( \begin{array}{ccc}
	0 & 0 & 0 \\
	0 & 0 & 0 \\
       	0 & 0 & 1 \\
         \end{array} \right) {U^v}^{\dagger} \nonumber  \\
M_{21} & = & U^v  \left( \begin{array}{ccc}
	0 & 0 & 0 \\
	0 & 1 & 0 \\
       	0 & 0 & 0 \\
         \end{array} \right) {U^v}^{\dagger} \nonumber  \\
\end{eqnarray}

Matter effects can be included by adding 
$A(r)$, to the $e-e$ element of $M_v^2$ where 
\begin{equation}
A  =  2 \sqrt{2} \left( \frac {G Y_{e}} {m_n} \right) \rho E .   
\label{eq:defA}
\end{equation}
In the above equation
$m_n$ is the mass of the nucleon,  $Y_{e}$ the number of electrons
per nucleon in the matter which is $\approx  \frac {1} {2} $, and $\rho$ is
the density of matter in gm/cc. $A$ can be written as
\[ A = 0.76 \times 10^{-4} \rho \times E . \]
$A$ is in $eV^2$, if $E$ is expressed in $GeV$.
In the atmospheric neutrino case most of the trajectories are through the
mantle of the earth. We take the mantle density to be a constant of value
$5$ gm/cc.

The matter
corrected $({\rm mass})^2$ matrix in the flavour basis is
\begin{equation}
M_m^2 = \Delta_{31} M_{31} + \Delta_{21} M_{21} + A M_A,
\label{eq:mmsq}
\end{equation}
where
\begin{equation}
M_A = \left( \begin{array}{ccc} 1 & 0 & 0 \\ 0 & 0 & 0 \\ 0 & 0 & 0 \\
       \end{array} \right).
\end{equation}

Now  $\Delta_{31} \gg \Delta_{21}$. 
Also for a typical neutrino energy in the multi-GeV range of about $3$ GeV
$\Delta_{21} \sim A$. Thus we work in an situation
where $\Delta_{21}, A \ll \Delta_{31}$.

In this approximation, to the zeroth order, both the matter term
and the term proportional to $\Delta_{21}$ can be neglected in
eq.~(\ref{eq:mmsq}). Then $M_m^2 = \Delta_{31} M_{31}$, whose
eigenvalues and eigenvectors are
\begin{eqnarray}
0 & ; & |1 \rangle = \left( \begin{array}{c} U_{e 1} \\ U_{\mu 1} \\ U_{\tau 1}
	\end{array} \right), \nonumber \\
0 & ; & |2 \rangle = \left( \begin{array}{c} U_{e 2} \\ U_{\mu 2} \\ U_{\tau 2}
	\end{array} \right), \nonumber \\
\Delta_{31} & ; & |3 \rangle = \left( \begin{array}{c} U_{e 3} \\ U_{\mu 3} \\ U_{\tau 3}
	\end{array} \right).
\end{eqnarray}
We treat $A M_A + \Delta_{21} M_{21}$ as perturbation to the
dominant term in $M_m^2$ and carry out degenerate perturbation
theory. 
We first define two new states as follows
\begin{eqnarray}
|1^{'}  \rangle &=& 
\alpha |1 \rangle + \beta |2 \rangle \\    
|2^{'}  \rangle &=& 
- \beta |1 \rangle + \alpha |2 \rangle.    
\end{eqnarray}
$\alpha$ and $\beta$ are determined by the conditions
\begin{equation}
\alpha^2 + \beta^2 = 1, ~~~{\rm and}~~~ 
\langle 1^{'} |H^{'}|2^{'} \rangle = 0, \label{eq:cond} 
\end{equation}
where
$H^{'} = A M_A + \Delta_{21} M_{21}$. 
The condition eq.~(\ref{eq:cond}) leads to the following equation
\begin{equation}
\frac {\alpha \sqrt{1 - \alpha^2}} {1 - 2 \alpha^2} = 
\frac {U_{e 1} U_{e 2}} {(\Delta_{21}/A) +(U_{e 2}^2-U_{e 1}^2)} 
\label{eq:alpha}
\end{equation}
Using the explicit form of the vacuum mixing matrix given in 
eqn.~(\ref{eq:defUv}) 
eq.~(\ref{eq:alpha}) can be written in terms of mixing angles as 
\begin{equation}
\frac {\alpha \sqrt{1 - \alpha^2}} {1 - 2 \alpha^2} = 
\frac {\frac{1}{2} \cos^2 \phi \sin 2 \omega} {(\Delta_{21}/A) 
-(\cos^2 \phi \cos 2 \omega)} 
\end{equation}
In the SB scheme $\omega$ is close to maximal mixing and $\phi$ is
small so for this scheme we get
\begin{equation}
\frac {\alpha \sqrt{1 - \alpha^2}} {1 - 2 \alpha^2} \simeq 
\frac {A} {2 \Delta_{21}} = k (say). \label{eq:defk} 
\end{equation}
From the above equation we get
\begin{eqnarray}
\alpha^2 &=& \frac{1}{2} \left[1 + \frac{1}{\sqrt{1+4 k^2}} \right] 
\label{eq:defaplsq} \\
\beta^2 &=& \frac{1}{2} \left[1 - \frac{1}{\sqrt{1+4 k^2}} \right]. 
\label{eq:defbetsq} 
\end{eqnarray}
For later use we also note that
\begin{eqnarray}
\alpha  \beta &=&  \frac{k} {\sqrt{1+4 k^2}}  \nonumber  \\
\alpha^2 - \beta^2 &=&  \frac{1} {\sqrt{1+4 k^2}}.  \nonumber  
\end{eqnarray}
The matter dependent mass eigenvalues are now given by
\begin{eqnarray}
m_1^2 & = & \beta^2 \Delta_{21} + A(\alpha^2 U_{e 1}^2 + \beta^2 U_{e 2}^2
	     + 2 \alpha \beta U_{e 1} U_{e 2}), \nonumber \\
m_2^2 & = & \alpha^2 \Delta_{21} + A(\alpha^2 U_{e 2}^2 + \beta^2 U_{e 1}^2
	     - 2 \alpha \beta U_{e 1} U_{e 2}), \nonumber \\
m_3^2 & = & \Delta_{31} + A U_{e 3}^2  \simeq \Delta_{31}.
\end{eqnarray}
Hence we get the mass squared differences in matter
\begin{eqnarray}
\Delta_{21}^{m}  & = & \frac{1}{\sqrt{1+4 k^2}} [\Delta_{21}- A \cos^2 \phi] , \nonumber \\
\Delta_{31}^{m}  & \simeq & \Delta_{31} , \nonumber \\
\Delta_{32}^{m}  & \simeq & \Delta_{31} 
\end{eqnarray}
By computing the first order correction to the old states one can derive the
mixing angles in matter. It is straightforward to show that two of the mixing
angles $\phi$ and $\psi$ are unaffected by matter (corrections of order
$(A/\Delta_{31}$ and are negligible). This was also 
demonstrated in \cite{cgk}.
The (12) mixing angle $\omega$ is strongly modified by matter effects and is
given by
\begin{equation}
\sin \omega_{m} = -\beta \cos \omega + \alpha \sin \omega
\end{equation}
Since $\omega$ is constrained to be close to $45$ degrees we get
\begin{eqnarray}
\sin \omega_{m} &=& \frac{1}{\sqrt{2}} (\alpha - \beta) \nonumber \\
&=& \frac{1}{\sqrt{2}}\left (\left[1 - \frac{\frac{A} {\Delta_{21}}}
{\sqrt{1+(\frac{A}{\Delta_{21}})^2}} \right]\right)^
{\frac{1}{2}} \label{eq:defomm}
\end{eqnarray}
Let us estimate the quantity $(A/\Delta_{21})$.
Using the definition of $A$ we get
\begin{equation}
\frac{A}{\Delta_{21}} = \frac{0.76 \times 10^{-4}\times 
5 \times E}{x \times 10^{-4}}, 
\label{eq:ratio}
\end{equation}
where we set $\rho = 5$ gm/cc, earth mantle density.
The quantity $x$ in the SB scheme typically is between 2 to 5.
So one sees that $(A/\Delta_{21})
 \simeq E$ where $E$ is the neutrino
energy in GeV. This implies
\begin{equation}
\sin \omega_{m} = \frac{1}{\sqrt{2}}\left (\left[1 - \frac{E}
{\sqrt{1+E^2}} \right]\right)^
{\frac{1}{2}} \label{eq:defommmgn}
\end{equation}
Now for typical multi-GeV events, $E \geq 2 GeV$. This in turn implies
\begin{equation}
\frac{E}{\sqrt{1+E^2}} \simeq 1 
\end{equation}
From eq.~(\ref{eq:defommmgn}) we see that $\sin \omega_{m} \rightarrow 0$.
In the case of two flavor mixing, this effect is known previously 
\cite{fogli1}. If the vacuum mixing angle is maximal, then the matter 
depedent mixing angle goes to zero at high energies. Here we have 
demonstrated that the same effect occurs in three flavor oscillations
also, in the limit of perturbation theory being valid. 
$\sin \omega_m = 0$ implies $U_{e2}^m = 0$ and hence
the (12) scale in matter, $\Delta_{21}^{m}$,
decouples from the oscillation probabilities involving
electron neutrinos. Below we show that the same thing happens
for electron anti-neutrinos also. 
The $\nu_\mu \leftrightarrow \nu_e$
oscillation probability in matter becomes
\begin{eqnarray}
P^m_{\mu e} &=& 4 U^m_{e1} U^m_{e3} U^m_{\mu 1} U^m_{\mu 3} 
\sin^2 \left(1.27 \frac{d \ \Delta_{31}}{E} \right) \nonumber \\
&=& \frac{1}{2} \sin^2 2 \phi \sin^2 \psi \label{eq:pmuen}
\end{eqnarray}
as the rapidly oscillating scale $\Delta_{31}$ will average out to half. 
This transition probability
is at most a few percent because of the smallness of $\phi$.
Let us now consider the case of antineutrinos. 
Since for antineutrinos $A \rightarrow -A$ we get 
\begin{equation}
\sin \omega_{m} = \frac{1}{\sqrt{2}}\left (\left[1 + \frac{E}
{\sqrt{1+E^2}} \right]\right)^
{\frac{1}{2}} \label{eq:defommmgan}
\end{equation}
In this case for typical multi-GeV energies we see
from eq.~(\ref{eq:defommmgan}) that $\sin \omega_{m} 
\rightarrow \frac{\pi}{2} $.
This implies $U_{e 1}^{m} = 0$. The $\bar{\nu}_\mu
\rightarrow \bar{\nu}_e$ oscillation probability becomes 
\begin{eqnarray}
P^m_{\bar{\mu} \bar{e}} &=& 4 U^m_{e2} U^m_{e3} U^m_{\mu 2} U^m_{\mu 3} 
\sin^2 \left(1.27 \frac{d \ \Delta_{31}}{E} \right) \nonumber \\
&=& \frac{1}{2} \sin^2 2 \phi \sin^2 \psi \label{eq:pmuean}
\end{eqnarray}
which is again a few percent.

Let us now contrast eq.~(\ref{eq:pmuen}) and 
eq.~(\ref{eq:pmuean}) with the expression
for the  $\nu_\mu \leftrightarrow \nu_e$ 
vacuum oscillation probability in the SB 
scheme. This is given by
\begin{eqnarray}
P^v_{\bar{\mu} \bar{e}} = P^v_{\mu e} & = &  
           - 4 U_{e1} U_{e2} U_{\mu 1} U_{\mu 2} 
\sin^2 \left(1.27 \frac{d \ \Delta_{21}}{E} \right) \nonumber \\ 
          & &  - 2 U_{e1} U_{e3} U_{\mu 1} U_{\mu 3} 
           - 2 U_{e2} U_{e3} U_{\mu 2} U_{\mu 3},
\label{eq:pmuev}
\end{eqnarray}
where we have set the rapid oscillating term due to $\Delta_{31}$
equal to $1/2$.
In the SB scheme, $\omega \simeq \pi/4$ and hence
the magnitudes of $U_{e 1}$, $U_{e 2}$,
$U_{\mu 1}$ and $U_{\mu 2}$ are $0.6$ or more.
Because of this, one gets large 
$\nu_\mu \leftrightarrow \nu_e$ transitions driven by  
$\Delta_{21}$, for large distances of travel 
(i.e. for upward going neutrinos). 

Let us now look at the survival probabilities for 
electron neutrinos and anti-neutrinos. By setting
$\omega_m = 0$ for neutrinos and $\omega_m = \pi/2$
for anti-neutrinos, we obtain the matter dependent
survival probabilities to be 
\begin{equation}
P^m_{ee} = P^m_{\bar{e} \bar{e}} =    
1 - \frac{1}{2} \sin^2 2 \phi \simeq 1. \label{eq:peem}
\end{equation}
We should constrast these with the vacuum survival probablilities
\begin{eqnarray}
P^v_{\bar{e} \bar{e}} = P^v_{e e} & = & 1 
           - 4 U_{e1}^2 U_{e2}^2 
\sin^2 \left(1.27 \frac{d \ \Delta_{21}}{E} \right) \nonumber \\ 
          & &  - 2 U_{e1}^2 U_{e3}^2  
           - 2 U_{e2}^2 U_{e3}^2  
\label{eq:peev}.
\end{eqnarray}
The last two terms in the above equation are proportional $\sin^2 \phi$
and are negligible. However the term containing $\Delta_{21}$ has a 
large coefficient and can cause significant oscillations for upward
going events. 

Similarly, matter depdendent survival probability for muon neutrinos 
(anti-neutrinos) are obtained by setting $\omega_m = 0 (\pi/2)$. They
turn out to be 
\begin{eqnarray}
P^m_{\mu \mu} & = & 1 - 4 \sin^2 2 \psi \sin^2 \phi  
\sin^2 \left(1.27 \frac{d \ \Delta^m_{21}}{E} \right) \nonumber \\ 
& & - 2 \cos^2 \phi \sin^2 \psi (1 - \cos^2 \phi \sin^2 \psi)  
\label{eq:pmumum}
\end{eqnarray}
The anti-neutrino survival probability has a similar expression, except
that $\Delta^m_{21}$ for anti-neutrinos is different from that of
neutrinos. However, in both cases, the zenith angle dependence is very
weak because the oscillations driven by $\Delta^m_{21}$ have a 
coefficient proportional to $\sin^2 \phi$ and hence are very small. 
In case of vacuum survival probabilities, there is siginificant 
zenith angle dependence coming from the oscillations of $\Delta_{21}$
whose coefficient $4 U_{\mu 1}^2 U_{\mu 2}^2$ is quite large.

Using the oscillation probabilities derived above, let us calculate
the zenith angle distributions of electron and 
muon events in atmospheric neutrinos.
They are given by
\begin{eqnarray}
N_\mu & = & \Phi_\mu P_{\mu \mu} + \Phi_e P_{e \mu} + \Phi_{\bar{\mu}}
P_{\bar{\mu} \bar{\mu}} + \Phi_{\bar{e}} P_{\bar{e} \bar{\mu}} \nonumber \\ 
N_e & = & \Phi_\mu P_{\mu e} + \Phi_e P_{e e} + \Phi_{\bar{\mu}}
P_{\bar{\mu} \bar{e}} + \Phi_{\bar{e}} P_{\bar{e} \bar{e}},
\end{eqnarray}  
where the products are to be understood as convolutions over energy.
To include matter effects, the matter dependent probabilities should
be used in the above equations.
The zenith angle dependence of vacuum probabilities, coming from
the oscillations driven by $\Delta_{21}$, gives the required zenith 
angle dependence to upward going muon events for multi-GeV energies.
For downward going muons, vacuum oscillations in SB scheme predict
larger suppression compared to what is observed. In case of electron
events, the vaccum oscillation prediction for downward going events
is in agreement with data but for upward going events, a huge excess
is predicted, which again occurs due to large $\nu_\mu \leftrightarrow
\nu_e$ oscillations driven by $\Delta_{21}$. When matter effects are
included, we have seen that $\Delta^m_{21}$  either decouples, as in 
$P_{ee}, P_{\bar{e} \bar{e}}, P_{\mu e}$ and $P_{\bar{\mu} \bar{e}}$ 
or has a very small coefficient, as in $P_{\mu \mu}$ and $P_{\bar{\mu}
\bar{\mu}}$. Hence, with the inclusion of the matter effects, the 
zenith angle distribution for both electron and muon event
distrutions become 
very flat. This is desirable from the point view of electron data
but the flat distribution strongly disagrees with the muon data. Since
the error bars in muon data are much smaller than those in electron
data, getting a better fit to electron data at the expense of muon 
data, worsens the overall fit of the data. Thus we see that inclusion
of matter effects in SB scheme worsen what was already not a good fit. 

We see from fig. (3) of ref.\cite{fogli} that the matter corrected
oscillations of SB scheme predict a mild zenith
angle distribution for the upward going multi-GeV electron events.
Whereas the upward going muon events are predicted to have a flat
distribution. These two features can also be explained in terms of 
our calculation. For electron events, the visible energy is the same
as the neutrino energy ($\geq 1.33$ GeV). For energy range of
$1.33-2$ GeV, at the beginning of multi-GeV events, $\omega_m
\neq 0$ for neutrinos (and $\omega_m \neq \pi/2$ for 
anti-neutrinos). Hence in this energy range, $\Delta_{21}^m$ does
not quite decouple. Since the flux is higher at lower energy, these
neutrinos lead to more events and hence we see some zenith angle
dependence of upgoing multi-GeV electron events. In case of muons, 
for a visible energy of $1.33$ GeV, the neutrino energy is greater
than $1.5$ GeV and hence the decoupling of $\Delta_{21}^m$ is 
better because of the higher neutrino energy threshold. 

For the through going muon events the neutrino energy is 
much higher than typical
multi-GeV energies. So again  
\begin{equation}
\frac{E}{\sqrt{1+E^2}} \simeq 1 
\end{equation}
and the discussions of the previous section apply and 
we will get a flat distribution
for the muon events when matter effects are included.
which is seen in the last panel of fig.(4) of \cite{fogli} 

Lastly we discuss the sub-GeV events. For these events
$A <  \Delta_{21}$. The parameter $k = (A/2 \Delta_{21})$ is 
about $15 \%$. For most of the energy range, the 
matter effects are negligible and one has essentially
vacuum oscillations. This can also be seen
in the first two panels of fig. (4) of ref.\cite{fogli}.

\section{Summary and Discussion}
We did an analytical calculation of matter effects in a scheme
where all the three evidences for neutrino oscillations are 
explained in terms of three active flavor oscillations. Our simple 
calculation reproduced most of the features of the matter effects
in this scheme, which were numerically investigated earlier.
Matter effects bring the predictions of this scheme for electron
events closer to the data and the fit to the electron data is 
improved. But they wipe out the zenith angle dependence of
multi-GeV muon events, which is in contradiction to the data.
Since the error bars of muon data are much smaller than those
of electron data, the deviation away from the data makes the
overall fit, with matter effects, to the data worse compared  
to the already bad fit of vacuum oscillation predictions of this
scheme. 

Finally we wish to make a rather interesting comment. If in 
future these kind of schemes become favored phenomenologically,
the main motivation for buildging a muon storage ring based 
neutrino factories is lost. At such factories, the neutrino
energies are in the range $10-50$ GeV. We have seen that at 
such large energies, due to earth matter effects, $\omega_m
\simeq 0$ for neutrinos and $\omega_m \simeq \pi/2$ for 
anit-neutrinos. In either case, the sub-dominant scale 
$\Delta^m_{21}$ decouples from oscillations and CP Violation
in lepton sector will be unmeasurably small in a neutrino 
factory environment. In such a case, the best option
to observe CP violation then will be a conventional superbeam
with a short baseline (of about 100 km) \cite{cbh}.
 
\medskip

{\bf Acknowledgements} : We wish to thank Sandhya Choubey, 
Srubabati Goswami and Kamales Kar for
discussions and providing us with numerical results of the 
SB scheme.

\end{document}